\documentclass{article}
\usepackage{arxiv}
\usepackage{float}
\usepackage{hhline}
\usepackage[utf8]{inputenc} 
\usepackage[T1]{fontenc}    
\usepackage{url}            
\usepackage{booktabs}       
\usepackage{amsfonts}       
\usepackage{nicefrac}       
\usepackage{microtype}      
\usepackage{lipsum}
\usepackage{graphicx}
\graphicspath{ {./images/} }

\title{Hierarchical Severity Staging of Anterior Cruciate Ligament Injuries using Deep Learning with MRI Images}

\usepackage{authblk}

\author[1,2,*]{Nikan K Namiri}
\author[1,2]{Io Flament}
\author[1,2]{Bruno Astuto}
\author[1,2]{Rutwik Shah}
\author[1,2]{Radhika Tibrewala}
\author[1,2]{Francesco Caliva}
\author[1,2]{Thomas M Link}
\author[1,2]{Valentina Pedoia}
\author[1,2]{Sharmila Majumdar}

\affil[1]{Department of Radiology and Biomedical Imaging, University of California, San Francisco, San Francisco, CA, USA}
\affil[2]{Center for Intelligent Imaging, University of California, San Francisco, San Francisco, CA, USA}
\affil[*]{Corresponding Author: Nikan K Namiri, nikan.namiri@ucsf.edu}

\begin{document}
\maketitle
\noindent \textbf{Summary Statement}:  This deep learning pipeline may lend towards diagnostic worklist prioritization, standardization, and generalizability in assessing anterior cruciate ligament lesions, in addition to point-of-care communication with patients by non-experts.\par
\noindent \textbf{Key Results}:\par
\begin{itemize}
	\item {Two convolutional neural networks (CNNs), each with respective two- and three-dimensional (2D and 3D) convolutional kernels, achieved a high overall accuracy of 92$\%$  (233/254) and 89$\%$  (225/254), respectively, and each a weighted Cohen’s kappa of 0.83 for ACL severity staging.}

	\item {All reconstructed ACLs were correctly classified by the 2D model (sensitivity of 100$\%$  (30/30) and specificity of 100$\%$  (224/224)), while the 3D CNN displayed similar performance (sensitivity of 97$\%$  (29/30) and specificity of 100$\%$  (223/224)).}

	\item {Sensitivity and specificity of the 2D and 3D models for intact classification were 93$\%$  (188/203) and 90$\%$  (46/51), and 89$\%$  (180/203) and 88$\%$  (45/51), respectively.}
\end{itemize}
\noindent \textbf{Note:} This work has been submitted to Radiology: Artificial Intelligence for possible publication. Copyright may be transferred without notice, after which this version may no longer be accessible. \par

\newpage

\begin{abstract}
\textbf{Purpose}: To evaluate the diagnostic utility of two convolutional neural networks (CNNs) for severity staging of anterior cruciate ligament (ACL) injuries. \newline\textbf{{Materials and Methods}}: This retrospective analysis was conducted on 1243 knee MR images (1008 intact, 18 partially torn, 77 fully torn, and 140 reconstructed ACLs) from 224 patients (age 47 $ \pm $  14 years, 54$\%$  women) acquired between 2011 and 2014. The radiologists used a modified scoring metric. To classify ACL injuries with deep learning, two types of CNNs were used, one with three-dimensional (3D) and the other with two-dimensional (2D) convolutional kernels. Performance metrics included sensitivity, specificity, weighted Cohen’s kappa, and overall accuracy, followed by McNemar’s test to compare the CNNs performance. \newline\textbf{Results}: The overall accuracy and weighted Cohen’s kappa reported for ACL injury classification were higher using the 2D CNN (accuracy: 92$\%$  (233/254) and kappa: 0.83) than the 3D CNN (accuracy: 89$\%$  (225/254) and kappa: 0.83) (\textit{P} = .27). The 2D CNN and 3D CNN performed similarly in classifying intact ACLs (2D CNN: 93$\%$  (188/203) sensitivity and 90$\%$  (46/51) specificity; 3D CNN: 89$\%$  (180/203) sensitivity and 88$\%$  (45/51) specificity). Classification of full tears by both networks were also comparable (2D CNN: 82$\%$  (14/17) sensitivity and 94$\%$  (222/237) specificity; 3D CNN: 76$\%$  (13/17) sensitivity and 100$\%$  (236/237) specificity). The 2D CNN classified all reconstructed ACLs correctly. \newline\textbf{Conclusion}: 2D and 3D CNNs applied to ACL lesion classification had high sensitivity and specificity, suggesting that these networks could be used to help grade ACL injuries by non-experts.  
\end{abstract}


\section{Introduction}
The anterior cruciate ligament (ACL) is the most commonly injured ligament in the knee (1,2). ACL injuries\  increase the risk of developing post-traumatic knee osteoarthritis and total knee replacement (3–6). MRI is currently the most effective imaging modality for distinguishing structural properties of the ACL in relation to adjacent musculoskeletal structures (7–10). Several multi-grading scoring systems have been developed to standardize reporting of knee joint abnormalities using MRI (11,12). The Whole-Organ Magnetic Resonance Imaging Scale (WORMS) is one of the most widely used semiquantitative MRI scoring system for knee osteoarthritis assessment (13–15) and considers a number of chondral, bony, and ligamentous compartments in the knee (15,16). The Anterior Cruciate Ligament OsteoArthritis Score (ACLOAS) has been shown to offer increased longitudinal and cross-sectional reliability of ACL staging (17). The aforementioned grading metrics are susceptible to inter-rater variability, especially pertaining to torn fibers and mucoid degeneration (10,18).

Deep learning methods have recently shown potential to serve as an aid for clinicians with limited time or experience in osteoarthritis grading of the knee menisci and cartilage (9). Four other applications of deep learning have also resulted in binary models for distinguishing an intact ACL from a fully torn ACL (19–22). These neural networks possess large domains for learning and are versatile for new tasks using pre-trained weights, producing inferences in seconds. The specific mode of learning depends on the architecture, but the most successful algorithms are known to learn shallow features and high-level features with many convolutions (23,24). However, deep learning has yet to be applied to predict multigrade, semiquantitative lesion severity for the ACL.

In this study, we propose a deep learning-based pipeline to isolate the ACL region of interest in the knee, detect ACL abnormalities, and stage lesion severity using three-dimensional (3D) and previously reported two-dimensional (2D) convolutions in MRIs. This is a proof-of-concept study to show that hierarchical image analysis methods work with 3D convolutional neural networks (CNNs) with the goal of hierarchical staging and comparison with the 2D network. Additionally, to the best of our knowledge this is the first instance of multiclass ACL severity staging using deep learning, in which reconstructed, fully torn, partially torn, and intact ACLs are graded in accordance with semi-quantitative scoring. This deep learning pipeline would lend towards standardization and generalizability in assessing ACL lesions for clinicians with limited time or those with limited experience reading knee MRIs.

\section{Materials and Methods}

\subsection{Patient Datasets}
This retrospective study was conducted according to regulations from the Committee for Human Research at all institutions, and all patients provided informed consent. Authors had full control of the data. A total of 1243 knee MRI studies (224 unique patients [mean age, 47 $ \pm $  14 years; mean body mass index, 24.58 $ \pm $  3.60 kg/m\textsuperscript{2}; 121 women], 1008 intact, 18 partially torn, 77 fully torn, and 140 reconstructed ACL images)were obtained from three prior research studies aimed to study joint degeneration in osteoarthritis and after ACL injury between 2011 and 2014. Patients were excluded based on concurrent use of an investigational drug, history of fracture, total knee replacement in the study knee, and any contraindications to MR. Patients were recruited in the osteoarthritis group if they reported knee pain, aching, or stiffness on most days per month during the past year, use of medication for knee pain on most days per month during the past year, or any possible radiologic sign of knee osteoarthritis (Kellgren-Lawrence > 0), and greater than 35 years of age. Patients were included in the control group if no knee pain or stiffness in either knee or use of medications for knee pain in the last year were reported, and if no radiologic evidence of osteoarthritis on either knee was noted (Kellgren-Lawrence = 0). Portions of this dataset were used in prior studies connected with grants NIH/NIAMS P50AR060752 and NIH/NIAMS R01AR046905. These studies had different aims such as relaxation time differences between loading and unloading in osteoarthritis patients, identifying cartilage change after ACL injury, and classifying knee meniscus and cartilage abnormalities (9).

Patients in the ACL study were enrolled at three sites: University of California, San Francisco (San Francisco, CA), Mayo Clinic (Rochester, MN), and Hospital for Special Surgery (New York City, NY). Patients (\textit{n }= 77) underwent anatomic single-bundle ACL reconstruction by board-certified, fellowship-trained orthopedic surgeons. Only soft tissue grafts were used, including the hamstring (either allograft or autograft) or the posterior tibialis (allograft). No special sequences were used for metal artifact suppression. Hamstring, patella tendon, and allograft ACL reconstructions typically do not cause metal artifacts along the intra-articular course of the graft and allow evaluation of ACL graft degeneration or retears. Moreover, metal artifacts related to ACL reconstruction are due to the use of metallic interference screws or Endobuttons, though many reconstructions typically make use of non-metallic interference screws and have no associated artifact. All patients underwent a standard postoperative rehabilitation protocol.

\subsection{MRI Acquisition}

In all imaging studies from all three sites, 3D fast spin-echo-Cube proton density-weighted sagittal oblique  sequences with the following parameters were used: repetition time / echo time = 1500 / 26.69 msec, field of view = 14 cm, matrix = 384 x 384, slice thickness = 0.5 mm, echo train length = 32, bandwidth = 50.0 kHz, number of excitations = 0.5, acquisition time = 10.5 min. All images were acquired with five 3T MRI scanners (GE Healthcare, Waukesha, WI) and 8 surface coils.

\subsection{Ground Truth Image Grading}

Between 2011 and 2014, five board-certified radiologists (each with over 5 years of training) graded a non-overlapping section of the dataset. The radiologists were blinded with respect to both number and type of lesion. An intra-observer agreement assessment was conducted by three additional board-certified radiologists currently involved with the study. All readers were trained by one senior musculoskeletal radiologist (T.M.L.) who read at least 20 imaging studies with each of the other two radiologists (radiology residents each with over 2 years of training) in two imaging sessions. During these readings, the WORMS and ACLOAS grading systems were explained and readers were asked to grade lesions of cartilage, menisci, bone marrow, ligaments, and synovium under supervision with direct feedback. This training was followed by independent assessment of 60 randomly chosen studies from the dataset. The ACL was graded using the original WORMS grading scale. The WORMS grade was then placed into four classes including intact (grade 0), partial tear (grade 1), full tear (grade 2), and reconstructed (grade 3) in order to generalize newer ACL scoring systems (ie ACLOAS) for use in the deep learning pipeline. The distribution of partial tears was small compared to that of full tears because the inclusion criteria in one of the studies was presence of fully torn ACL.

\subsection{Deep Learning Pipeline}

Our framework consists of a deep learning segmentation that categorizes the knee into 11 distinct anatomical components, followed by an image cropping to isolate the ACL, and a 3D CNN to classify lesion severity (9) (\textbf{Figure 1}). The segmentation and cropping for ACL localization are described in \textbf{Appendix E1 (supplement).} The CNN was developed through a hierarchical approach; specifically, three cascaded models were built to classify reconstructed ligaments, full tears, partial tears, and intact ACLs. We further compared the hierarchical approach to a single four class model with the same parameters, and saw superior performance using the hierarchical approach. The same hierarchical classification network was implemented with a 2D CNN for comparison with the 3D CNN (25).

\begin{figure}[!htb]

	\centering
    \includegraphics[width=6.5in,height=3.14in]{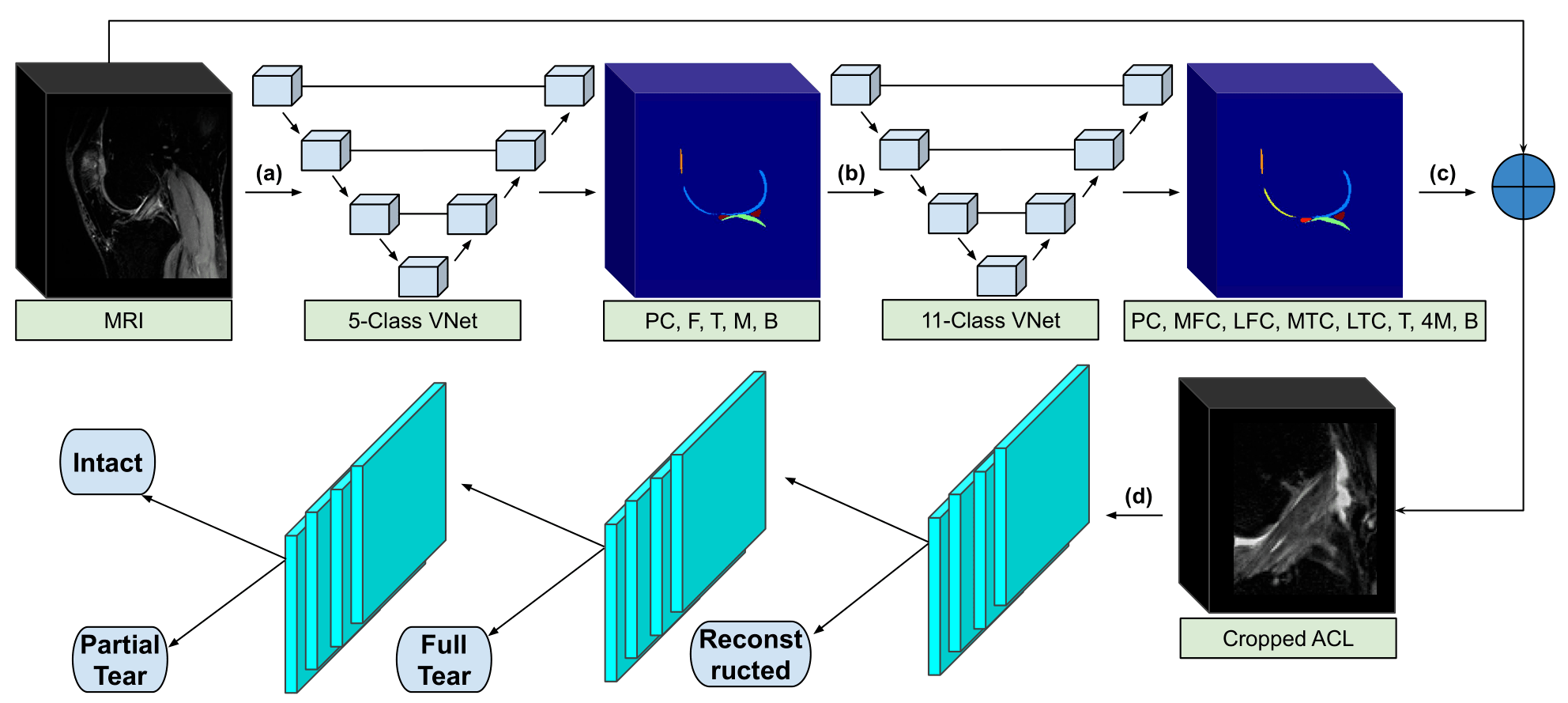}
    \caption{The segmentation and classification pipeline begins with \textbf{(a)} the input of a full MR volume into a three-dimensional VNet which segments the knee into patellar cartilage (PC), femur (F), tibia (T), meniscus (M), and background (B). \textbf{(b)} The 5-class segmentation is then input to a second VNet which further categorizes the knee into 11 compartments including patellar cartilage (PC), medial and lateral femoral and tibial condyles (MFC, LFC, MTC, LTC), and four meniscal horns (4M). \textbf{(c)} The 11-class segmentation is used to determine the ACL boundaries of the original input MRI. \textbf{(d)} The cropped ACL volume is input to three hierarchical convolutional neural networks (either two- or three-dimensional), which each detect reconstructed, fully torn, partially torn, and intact anterior cruciate ligaments.}

\end{figure}


\subsection{Classification of the 3D and 2D CNNs}

The 3D CNN was developed in Tensorflow (Google, Mountain View, CA), and the 2D CNN in Pytorch (Facebook, Menlo Park, CA).  All computations were performed on NVIDIA (Santa Clara, CA) GeForce GTX Titan X graphics processing units.

\textit{3D CNN}.—\  The cropped ACL volumes were input into a CNN consisting of 3D convolutional kernels (9). The network is built with six layers, including one skip connection after the first convolution, to preserve initial features and mitigate overfitting (\textbf{Figure 2}). Training was performed over 100 epochs with the following parameters: an Adam optimizer, a learning rate of 10\textsuperscript{-5}, empirically weighted cross-entropy loss to account for class imbalances, and a batch size of 8. Three-dimensional translations and zooming were applied to all classes for augmentation of the training set. Rotations were not applied in order to preserve ligament fiber angles for model learning.


\begin{figure}
	\centering
	\includegraphics[width=6.5in,height=1.45in]{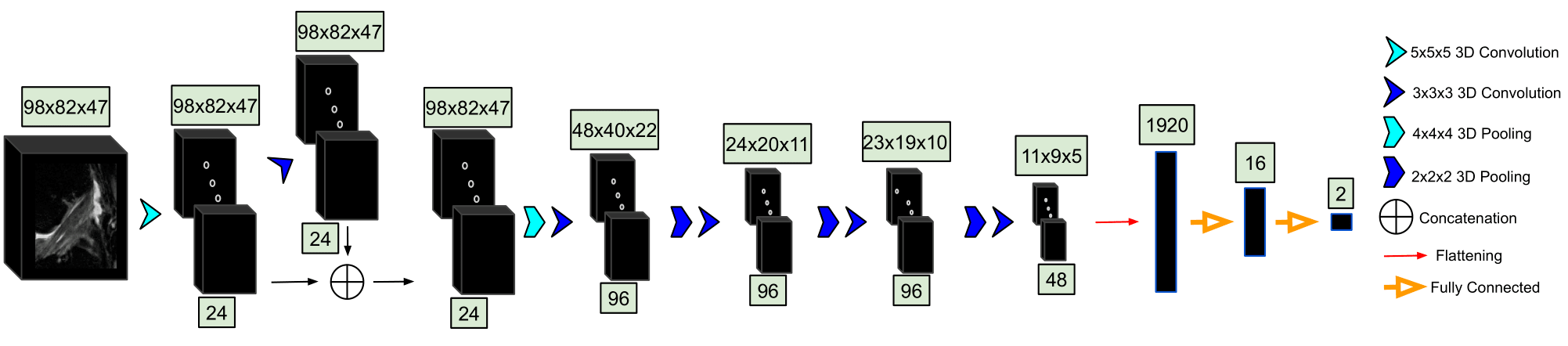}
	\caption{The MRI is input to the three-dimensional convolutional neural network. In the first layer, convolutional kernels are applied to the entire volume. The second layer contains a concatenation with the first, followed by four additional three-dimensional convolutions. The convolutional output is then flattened and input to two dense layers. The number beneath each set of blocks denotes the number of convolutions applied to the input; the number above represents the dimensions of the output after convolving the input.}
\end{figure}


\textit{2D CNN}.—Performance of the 3D CNN was compared with the MRNet, a 2D CNN (19). In the MRNet, each slice of the input 3D volume is passed through an AlexNet to extract features (25) (\textbf{Figure 3}).\  The MRNet was pre-trained on the ImageNet dataset, and additionally trained with the same training sets as the 3D CNN using transfer learning. The MRNet pools features within each slice and among all slices in the volume to produce a classification probability. The same image augmentations and loss functions as the 3D CNN were used to ensure correct comparison.


\begin{figure}[!htb]
    \centering
	\includegraphics[width=6.5in,height=3.16in]{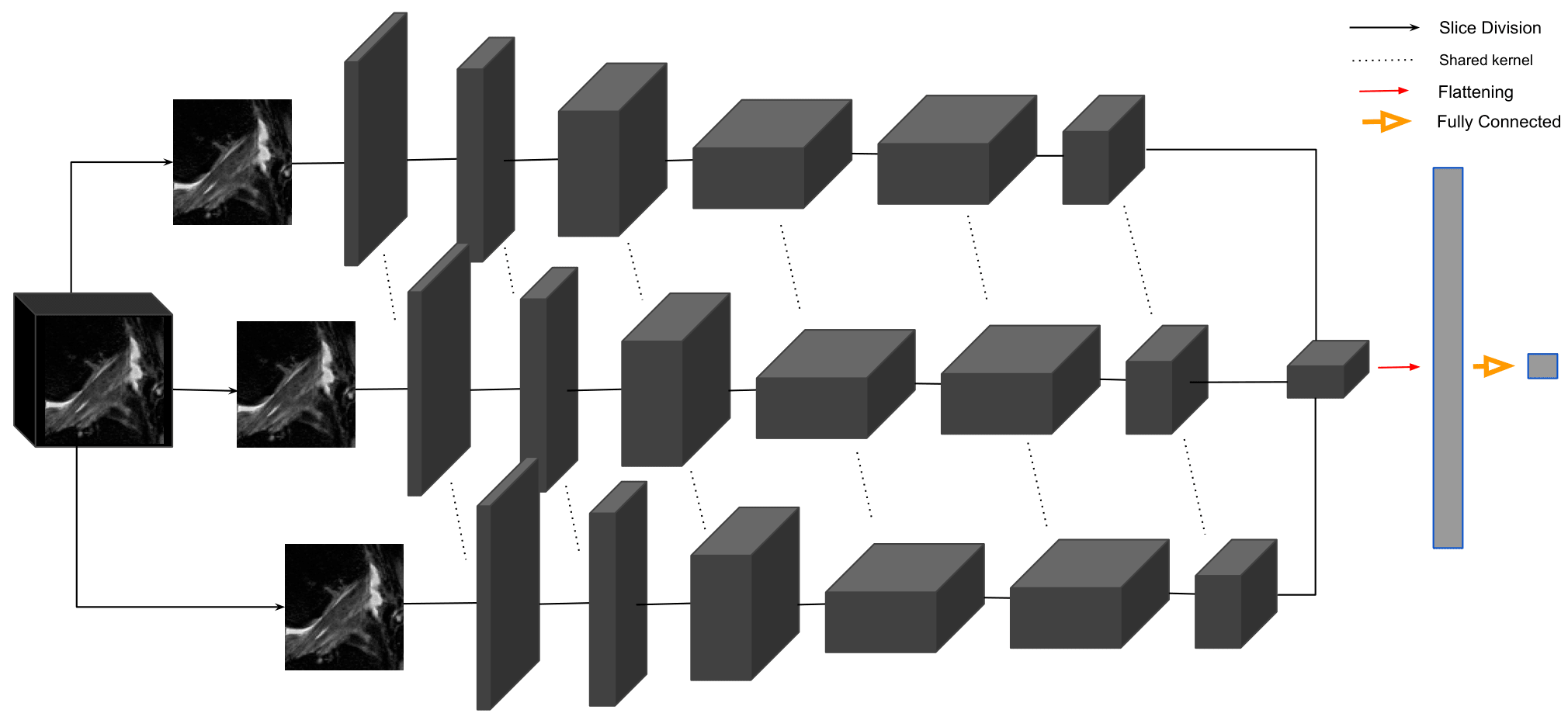}
	\caption{The two-dimensional convolutional neural network, MRNet, feeds each slice of an MRI into an AlexNet with shared parameters across each kernel. The weights are pooled as they pass into a final convolution and a subsequent fully dense flattening layer. The output of the dense layer is a single probability for the input MRI. The full parameters of the architecture are reported by Bien et al (19).}
\end{figure}


\subsection{Learning Strategy}
The deep learning classifier first differentiated the reconstructed ligaments (grade 3) from grades 0-2. The remaining studies were then analyzed for detecting full thickness tears (grade 2).\  Partial tear lesions (grade 1) were further classified apart from intact ligaments (grade 0). The total of 1243 images (1008 control, 235 injured) from 224 patients were split into training (70$\%$ ), validation (10$\%$ ), and test sets (20$\%$ ) for each grade, preserving the distributions of age, sex, and BMI. The studies in each split were from distinct, non-overlapping patients.

\subsection{Statistical Analysis}
The training set was used to train each of the CNNs with back propagation. The validation set served to evaluate model performance at each training epoch, and the testing set was blind to the model until after training to serve as final metric of performance. For each CNN, we report overall accuracy and linear-weighted Cohen’s kappa (26), as well as sensitivity and specificity for each severity score.  McNemar’s test determined statistical significance between the two classifiers for sensitivity, specificity, and overall accuracy; however, Fisher’s exact test was used instead if the number of patients with differing test results were less than 20 (27). Two-sample t-tests were used to compare training, validation, and test set demographics. Python v3.6.5 (Python Software Foundation) was used for all statistical analysis. \textit{P} values less than .05 were considered significant.

\section{Results}
The intra-reader agreement assessment for ACL grading resulted in linear weighted kappas of 0.66-0.78 among each pair of radiologists. A random sample of 17 ACL volumes resulted in a mean $ \pm $  standard deviation intersection over union of 0.89 $ \pm $  0.06. There were no statistically significant differences between patients in the training, validation, and test sets regarding age (\textit{P} = .24), BMI \textit{(P} = .17), or sex \textit{(P} = .85).

\subsection{Performance Metrics of the 2D and 3D CNNs}

The 3D CNN and 2D CNN had overall accuracies of 89$\%$  (225/254) and 92$\%$  (233/254) \textit{(P} = .27), respectively, and both CNNs had a linear-weighted kappa of 0.83 for ACL staging. \textbf{Table 1} displays the distribution of grades for model training, while \textbf{Table 2 }contains the total grades and demographics for the training, validation, and test sets. As seen in \textbf{Figure 4}, both models performed highest on reconstructed ACL classification. The sensitivities (2D CNN: 100$\%$  (30/30), 3D CNN: 97$\%$  (29/30)) and specificities (2D CNN: 100$\%$  (224/224), 3D CNN: 100$\%$  (224/224)) in reconstructed classification were not significantly different between the 2D CNN and 3D CNN (\textit{P }= 1.0, \textit{P} = 1.0). The 2D CNN demonstrated higher sensitivity and specificity in detecting intact ACLs (2D CNN: 93$\%$  (188/203) sensitivity and 90$\%$  (46/51) specificity; 3D CNN: 89$\%$  (180/203) sensitivity and 88$\%$  (45/51) sensitivity) (\textbf{Table 3}). The sensitivity of the 2D CNN in fully torn ACLs (82$\%$  (14/17) was similar to that of the 3D CNN (76$\%$  (13/17) (\textit{P} = 1.0), though the specificity of the 3D CNN (100$\%$  (236/237) was higher than that of the 2D CNN (94$\%$  (222/237) (\textit{P} < .001).


\begin{table}[!htb]
\caption{Distribution of Severity Gradings in Training Set and Groupings for each of the Hierarchical Neural Network Classifiers. Note.— Numbers within parentheses are image count and column-wise percentage. FT = full tear, I = intact, PT = partial tear, R = reconstructed.} 

 			\centering
\begin{tabular}{p{0.98in}|p{1.68in}|p{1.46in}|p{1.55in}}\hline 
\hline
\multicolumn{1}{|p{0.98in}}{{  \textbf{Classifier Type}}} & 
\multicolumn{1}{|p{1.68in}}{{  \textbf{Reconstructed}}} & 
\multicolumn{1}{|p{1.46in}}{{  \textbf{Full Tear}}} & 
\multicolumn{1}{|p{1.55in}|}{{  \textbf{Partial Tear}}} \\
\hhline{----}
\multicolumn{1}{|p{0.98in}}{{  \textbf{Negative Class}}} & 
\multicolumn{1}{|p{1.68in}}{{  FT + PT + I (88.9$\%$ , 766/862)}} & 
\multicolumn{1}{|p{1.46in}}{{  PT + I (93.2$\%$ , 714/766)}} & 
\multicolumn{1}{|p{1.55in}|}{{  I (98.3$\%$ , 702/714)}} \\
\hhline{----}
\multicolumn{1}{|p{0.98in}}{{  \textbf{Positive Class}}} & 
\multicolumn{1}{|p{1.68in}}{{  R (11.1$\%$ , 96/862)}} & 
\multicolumn{1}{|p{1.46in}}{{  FT (6.8$\%$ , 52/766)}} & 
\multicolumn{1}{|p{1.55in}|}{{  PT (1.7$\%$ ,12/714)}} \\
\hhline{----}

\end{tabular}
\end{table}



\begin{table}[!htb]
\caption{Distribution of ACL Gradings in 224 Patients. Note.— The mean $ \pm $  standard deviation is shown for age and BMI.}
\centering
\begin{tabular}{p{1.42in}p{1.42in}p{1.42in}p{1.42in}}
\hline
\multicolumn{1}{|p{1.42in}}{{   \textbf{Characteristic}}} & 
\multicolumn{1}{|p{1.42in}}{\centering {   \textbf{Training Set}}} & 
\multicolumn{1}{|p{1.42in}}{\centering {   \textbf{Validation Set}}} & 
\multicolumn{1}{|p{1.42in}|}{\centering {   \textbf{Test Set}}} \\
\hhline{----}
\multicolumn{1}{|p{1.42in}}{{   \textbf{Age, y}}} & 
\multicolumn{1}{|p{1.42in}}{\centering {   48.05 $ \pm $  12.84}} & 
\multicolumn{1}{|p{1.42in}}{\centering {   44.77 $ \pm $  16.31}} & 
\multicolumn{1}{|p{1.42in}|}{\centering {   42.98 $ \pm $  13.50}} \\
\hhline{----}
\multicolumn{1}{|p{1.42in}}{{   \textbf{BMI (kg/m\textsuperscript{2})}}} & 
\multicolumn{1}{|p{1.42in}}{\centering {   24.28 $ \pm $  3.52}} & 
\multicolumn{1}{|p{1.42in}}{\centering {   25.06 $ \pm $  3.95}} & 
\multicolumn{1}{|p{1.42in}|}{\centering {   25.18 $ \pm $  3.61}} \\
\hhline{----}
\multicolumn{1}{|p{1.42in}}{{   \textbf{Women}}} & 
\multicolumn{1}{|p{1.42in}}{\centering {   56.9$\%$  (83/146)}} & 
\multicolumn{1}{|p{1.42in}}{\centering {   50.0$\%$  (13/26)}} & 
\multicolumn{1}{|p{1.42in}|}{\centering {   48.1$\%$  (25/52)}} \\
\hhline{----}
\multicolumn{1}{|p{1.42in}}{{   \textbf{Intact}}} & 
\multicolumn{1}{|p{1.42in}}{\centering {   81.4$\%$  (702/862)}} & 
\multicolumn{1}{|p{1.42in}}{\centering {   81.1$\%$  (103/127)}} & 
\multicolumn{1}{|p{1.42in}|}{\centering {   79.9$\%$  (203/254)}} \\
\hhline{----}
\multicolumn{1}{|p{1.42in}}{{   \textbf{Partial Tear}}} & 
\multicolumn{1}{|p{1.42in}}{\centering {   1.4$\%$  (12/862)}} & 
\multicolumn{1}{|p{1.42in}}{\centering {    1.6$\%$  (2/127)}} & 
\multicolumn{1}{|p{1.42in}|}{\centering {   1.6$\%$  (4/254)}} \\
\hhline{----}
\multicolumn{1}{|p{1.42in}}{{   \textbf{Full Tear}}} & 
\multicolumn{1}{|p{1.42in}}{\centering {   6.0$\%$  (52/862)}} & 
\multicolumn{1}{|p{1.42in}}{\centering {   6.3$\%$  (8/127)}} & 
\multicolumn{1}{|p{1.42in}|}{\centering {   6.7$\%$  (17/254)}} \\
\hhline{----}
\multicolumn{1}{|p{1.42in}}{{   \textbf{Reconstructed}}} & 
\multicolumn{1}{|p{1.42in}}{\centering {   11.1$\%$  (96/862)}} & 
\multicolumn{1}{|p{1.42in}}{\centering {   11.0$\%$  (14/127)}} & 
\multicolumn{1}{|p{1.42in}|}{\centering {   11.8$\%$  (30/254)}} \\
\hhline{----}

\end{tabular}
 \end{table}
 

\begin{figure}[!htb]
	\centering
	\includegraphics[width=6.5in,height=2.58in]{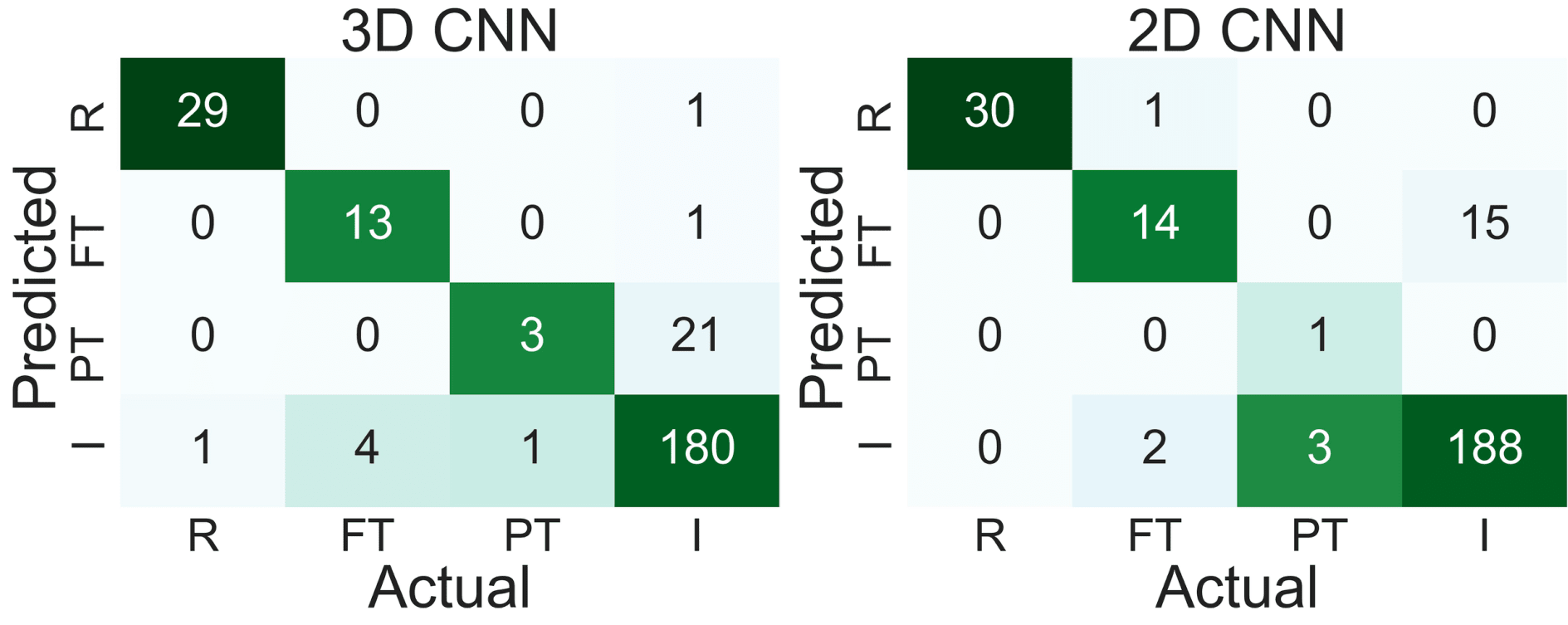}
	\caption{Confusion matrices for two- and three-dimensional convolutional neural networks of reconstructed (R), fully torn (FT), partially torn (PT), and intact (I) anterior cruciate ligaments.}
\end{figure}



\begin{table}[!htb]
\centering
\caption{Sensitivity and Specificity for 3D and 2D CNNs in Hierarchical Severity Staging of ACL Injuries.}
\begin{tabular}{p{0.98in}p{0.85in}p{0.85in}p{0.82in}p{0.84in}p{0.85in}p{0.85in}}
\hline
\multicolumn{1}{|p{0.98in}}{{    \textbf{Severity}}} & 
\multicolumn{1}{|p{0.85in}}{{    \textbf{3D Sensitivity ($\%$ )}}} & 
\multicolumn{1}{|p{0.85in}}{{    \textbf{2D Sensitivity ($\%$ )}}} & 
\multicolumn{1}{|p{0.82in}}{{    \textbf{\textit{P} Value}}} & 
\multicolumn{1}{|p{0.84in}}{{    \textbf{3D Specificity ($\%$ )}}} & 
\multicolumn{1}{|p{0.85in}}{{    \textbf{2D Specificity ($\%$ )}}} & 
\multicolumn{1}{|p{0.85in}|}{{    \textbf{\textit{P} Value}}} \\
\hhline{-------}
\multicolumn{1}{|p{0.98in}}{{    \textbf{Intact}}} & 
\multicolumn{1}{|p{0.85in}}{{    89 (180/203)}} & 
\multicolumn{1}{|p{0.85in}}{{    93 (188/203)}} & 
\multicolumn{1}{|p{0.82in}}{{    .22}} & 
\multicolumn{1}{|p{0.84in}}{{    88 (45/51)}} & 
\multicolumn{1}{|p{0.85in}}{{    90 (46/51)}} & 
\multicolumn{1}{|p{0.85in}|}{{    1.0}} \\
\hhline{-------}
\multicolumn{1}{|p{0.98in}}{{    \textbf{Partial Tear}}} & 
\multicolumn{1}{|p{0.85in}}{{    75 (3/4)}} & 
\multicolumn{1}{|p{0.85in}}{{    25 (1/4)}} & 
\multicolumn{1}{|p{0.82in}}{{    .49}} & 
\multicolumn{1}{|p{0.84in}}{{    92 (229/250)}} & 
\multicolumn{1}{|p{0.85in}}{{    100 (250/250)}} & 
\multicolumn{1}{|p{0.85in}|}{{   \textit{< }.001}} \\
\hhline{-------}
\multicolumn{1}{|p{0.98in}}{{    \textbf{Full Tear}}} & 
\multicolumn{1}{|p{0.85in}}{{    76 (13/17)}} & 
\multicolumn{1}{|p{0.85in}}{{    82 (14/17)}} & 
\multicolumn{1}{|p{0.82in}}{{    1.0}} & 
\multicolumn{1}{|p{0.84in}}{{    100 (236/237)}} & 
\multicolumn{1}{|p{0.85in}}{{    94 (222/237)}} & 
\multicolumn{1}{|p{0.85in}|}{{   \textit{< }.001}} \\
\hhline{-------}
\multicolumn{1}{|p{0.98in}}{{    \textbf{Reconstructed}}} & 
\multicolumn{1}{|p{0.85in}}{{    97 (29/30)}} & 
\multicolumn{1}{|p{0.85in}}{{    100 (30/30)}} & 
\multicolumn{1}{|p{0.82in}}{{    1.0}} & 
\multicolumn{1}{|p{0.84in}}{{    100 (223/224)}} & 
\multicolumn{1}{|p{0.85in}}{{    100 (224/224)}} & 
\multicolumn{1}{|p{0.85in}|}{{   1.0}} \\
\hhline{-------}

\end{tabular}
 \end{table}

\subsection{Examples of Correct and Incorrect Classifications}

\textbf{Figure} \textbf{5 }displays a knee with intact ACL that was input into the pipeline, followed by localization of the ACL and the corresponding saliency map generated by the model’s classification weighting. Saliency maps were generated from the 3D CNN’s rectified linear unit output in its last dense layer. This ACL was correctly classified and possesses a high intensity on the inferior portion of the ligament’s saliency. An incorrectly classified intact ACL, predicted to be partially torn, is seen in \textbf{Figure 6}.\textbf{ }The model placed a high intensity on a sagittal view with overlapping ACL and femur signal. The resulting saliency possessed large weight on a portion of the joint posterior to the ligament and also has speckles of noise further posterior.  The 3D CNN took less than one second to classify a single ACL that went through all three hierarchical classifiers.


\begin{figure}[!htb]
	\centering
	\includegraphics[width=6.5in,height=4.25in]{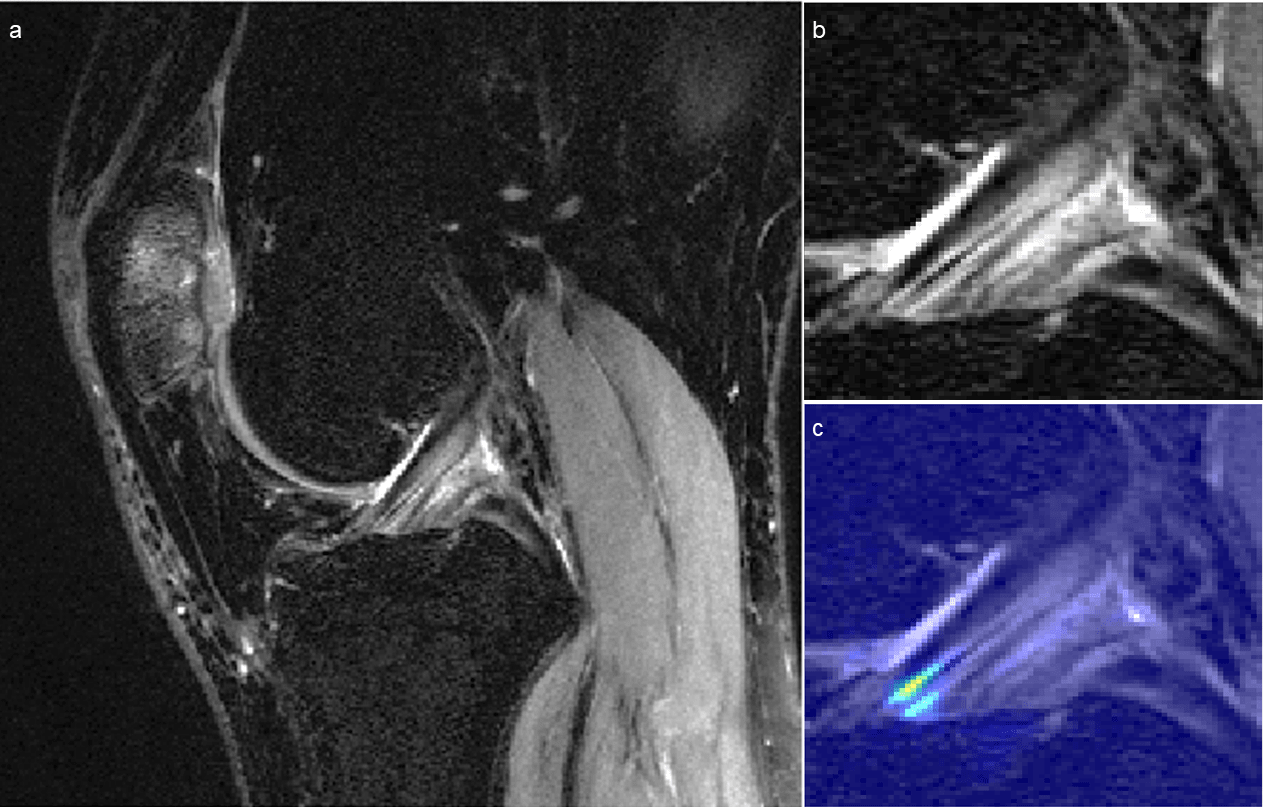}
	\caption{Sagittal views of \textbf{(a)} correctly classified knee with intact anterior cruciate ligament (ACL) and \textbf{(b)} its ACL localization with \textbf{(c)} overlaid saliency map. The saliency demonstrates the anterior-inferior portion of the ACL as high importance for model classification.}
\end{figure}



\begin{figure}[!htb]
	\centering
	\includegraphics[width=6.5in,height=4.25in]{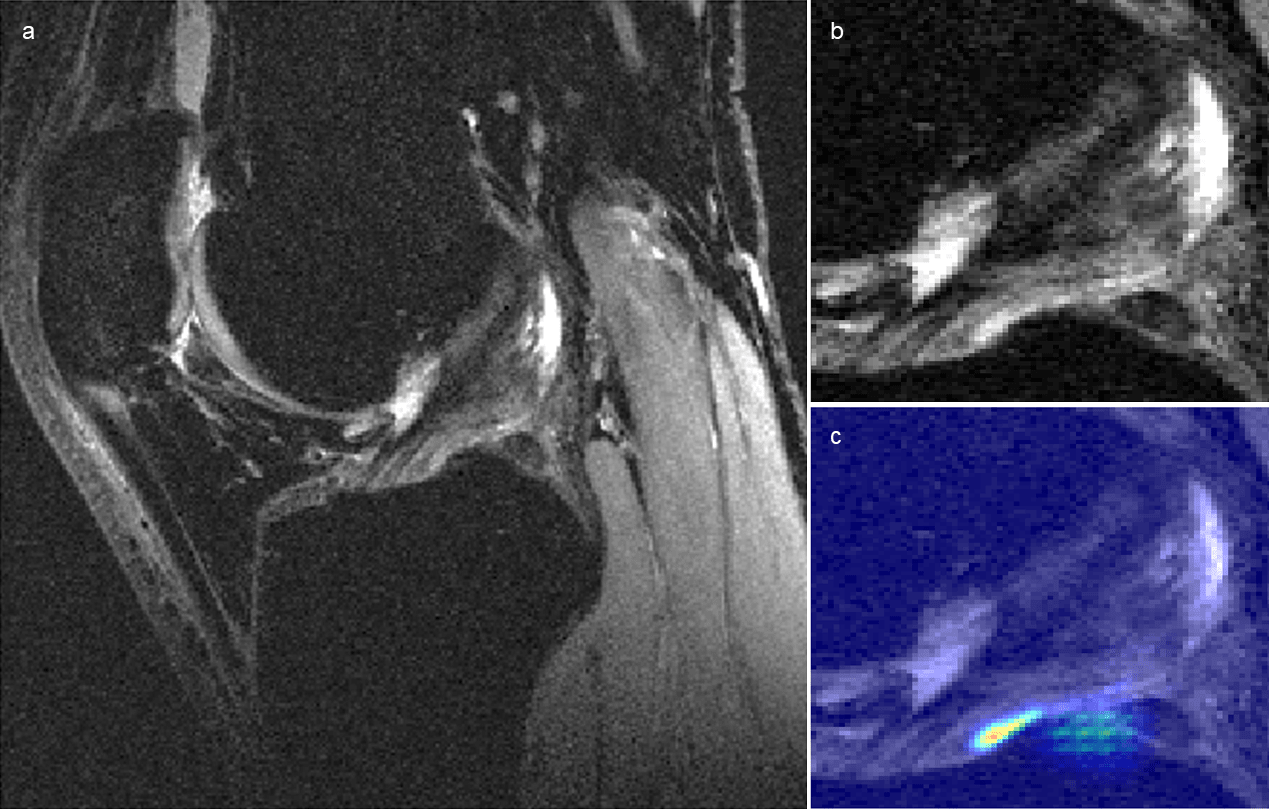}
	\caption{Sagittal views of \textbf{(a)} incorrectly classified knee with intact anterior cruciate ligament (ACL) and \textbf{(b)} its ACL localization with \textbf{(c)} overlaid saliency map. The model predicted a partial tear in this knee. The model misplaces a relatively high probability mapping on a slice with ACL that is obstructed by signal from femur. Additionally, the saliency intensity is posterior to the actual ligament; speckles of noise are also present in the inferior-posterior portion of the saliency. This intact ACL possessed focal fluid collection posterior to the ligament on a separate sagittal view, which may have led to the misclassification.}
\end{figure}


\section{Discussion}
In this work we present a fully-automated ACL segmentation and classification framework which provides hierarchical severity staging of the ACL using deep learning architectures. We compare the performances of a 3D and a 2D CNN in ACL lesion classification. A higher overall accuracy was observed with the 2D model.

Four groups have previously used deep learning frameworks for ACL lesion classification tasks, the first of which being the original MRNet by Bien et al (19). Their dataset consisted of 319 ACL tears from a total of 1,370 exams. Their MRNet displayed 97$\%$  sensitivity and 76$\%$  specificity for fully torn ACLs. The MRNet that we built demonstrated a higher specificity than sensitivity. The discrepancy may be due to difficulty in generalizing the MRNet to images from other institutions. Liu et al have approached the binary ACL tear classification task using three cascaded deep learning architectures\textit{ }(28–30). Their cascaded model achieved 96$\%$  in both sensitivity and specificity but was not statistically significant when compared to radiologist grading (20). In addition, this group used a relatively small number of images (350) for training, validation, and testing, which may have led to overfitting. Chang et al applied CNNs with residual blocks on 260 volumes in the coronal plane, which is a more difficult cross-section to use to grade the ACL (21). Most recently, Germann et al used CNNs to classify ACL tears in MRIs from 59 institutions, resulting in overall and in-house performance metrics lower than those of fellowship-trained faculty musculoskeletal radiologists (22). Compared to these four prior studies, our pipeline classifies ACLs using 3D convolutional kernels in a hierarchical sequence. We implemented a hierarchical classifier because the ACLs possess an ordered sequence of injury severity, increasing from intact to reconstructed. A stepwise approach beginning sequentially with the most severe classification can enhance accuracy because decisions are less complicated if they are binary.

MRIs are 3D volumes and 3D convolutional kernels can learn 3D features that 2D convolutions cannot. The ground truth we are using is on the patient level on all the 3D volumes, which is a scalable design because it does not require pixel or slice level annotation. Thus, supervised feature learning can occur exclusively in 3D, as opposed to 2D. However, 3D models are more complex, higher parameters spaces that are more likely to overfit with small sample sizes, unlike 2D models which have lower parameter space convolutional filters and typically perform well for general image classification problems. Using a 3D CNN did not outperform a 2D model, which may be because the 2D CNN utilized transfer learning from ImageNet, which is a dataset of 14 million 2D images that is not compatible for training a 3D CNN. The pre-trained 2D CNN without transfer learning would perform worse had we not trained it with our MRIs. However, our goal was to compare transfer learning in a 2D CNN with a 3D CNN possessing no transfer learning, in order to compare the benefits of 3D spatial relations with those of transfer learning.

Although we evaluate a hierarchical severity staging classifier, the partial tear class has little clinical relevance without subcategorization because partial tears denote a wide spectrum of injury, some of which require surgery and others which do not. Our limited sample size of partial tears did not allow for sub-stratification, but one of the primary end goals of this study is to classify subcategories of partial tears ranging from intact to fully torn. The limited number of partial tears in validation and testing sets should be considered when externalizing the partial tear results, as many more cases are needed to draw significant conclusions.

Our method contains other limitations beginning with the use of MRI as reference standard. The grades used for model training are dependent on subjective assessment by a radiologist. Using arthroscopy as an additional standard of comparison may improve the ground truth labels for the ACLs, increasing the model’s capability to learn accurately. However, the model learning is limited as early model building notably could not sub-classify mucoid degeneration within intact ligaments. Furthermore, our sample of patients was not balanced amongst all gradings, which we addressed by using a weighted cross entropy loss function during training for both 3D and 2D CNN models.  Another limitation of our study is that we split our dataset into training, validation, and testing sets according to patient. This may lead to correlations among multiple images from the same patient, which are non-independent observations. Accounting for such correlations in training neural networks is not common, as the majority of models are built using image augmentations to increase training capacity. Thus, even if each patient had solely one image, data augmentation would yet lend towards correlation discrepancies. For this reason, dividing by images without preserving patient splits would offer little correlation benefit for the model, which would instead display falsely elevated accuracy by inferring on the same patients used for training.

Deep learning can provide relatively fast classification and visualization of ACL lesions, which may facilitate clinical translation. Both the 2D and 3D architectures displayed a relatively high degree of sensitivity and specificity for intact, fully torn, and reconstructed ACLs, which may warrant clinical value of deep learning as a tool for standardizing and generalizing ACL severity staging for clinicians with limited experience with knee MRI.

\bibliographystyle{unsrt}  

\newpage
\section{References}

1. Spindler KP, Wright RW. Anterior cruciate ligament tear. N Engl J Med. Mass Medical Soc; 2008;359(20):2135–2142.
\newline
2. Johnston JT, Mandelbaum BR, Schub D, et al. Video analysis of anterior cruciate ligament tears in professional American football athletes. Am J Sports Med. SAGE Publications Sage CA: Los Angeles, CA; 2018;46(4):862–868.
\newline
3. Hunter DJ, Lohmander LS, Makovey J, et al. The effect of anterior cruciate ligament injury on bone curvature: exploratory analysis in the KANON trial. Osteoarthr Cartil. Elsevier; 2014;22(7):959–968.
\newline
4. Prodromos CC, Han Y, Rogowski J, Joyce B, Shi K. A meta-analysis of the incidence of anterior cruciate ligament tears as a function of gender, sport, and a knee injury–reduction regimen. Arthrosc J Arthrosc Relat Surg. Elsevier; 2007;23(12):1320–1325.
\newline
5. Brophy RH, Gill CS, Lyman S, Barnes RP, Rodeo SA, Warren RF. Effect of anterior cruciate ligament reconstruction and meniscectomy on length of career in National Football League athletes: a case control study. Am J Sports Med. Sage Publications; 2009;37(11):2102–2107.
\newline
6. Suter LG, Smith SR, Katz JN, et al. Projecting lifetime risk of symptomatic knee osteoarthritis and total knee replacement in individuals sustaining a complete anterior cruciate ligament tear in early adulthood. Arthritis Care Res (Hoboken). Wiley Online Library; 2017;69(2):201–208.
\newline
7. Shakoor D, Guermazi A, Kijowski R, et al. Cruciate ligament injuries of the knee: A meta‐analysis of the diagnostic performance of 3D MRI. J Magn Reson Imaging. Wiley Online Library; 2019.
\newline
8. Ai T, Zhang W, Priddy NK, Li X. Diagnostic performance of CUBE MRI sequences of the knee compared with conventional MRI. Clin Radiol. Elsevier; 2012;67(12):e58–e63.
\newline
9. Pedoia V, Norman B, Mehany SN, Bucknor MD, Link TM, Majumdar S. 3D convolutional neural networks for detection and severity staging of meniscus and PFJ cartilage morphological degenerative changes in osteoarthritis and anterior cruciate ligament subjects. J Magn Reson Imaging. Wiley Online Library; 2019;49(2):400–410.
\newline
10. Li K, Du J, Huang L-X, Ni L, Liu T, Yang H-L. The diagnostic accuracy of magnetic resonance imaging for anterior cruciate ligament injury in comparison to arthroscopy: a meta-analysis. Sci Rep. Nature Publishing Group; 2017;7(1):7583.
\newline
11. Hunter DJ, Guermazi A, Lo GH, et al. Evolution of semi-quantitative whole joint assessment of knee OA: MOAKS (MRI Osteoarthritis Knee Score). Osteoarthr Cartil. Elsevier; 2011;19(8):990–1002.
\newline
12. Brandt KD, Fife RS, Braunstein EM, Katz B. Radiographic grading of the severity of knee osteoarthritis: relation of the Kellgren and Lawrence grade to a grade based on joint space narrowing, and correlation with arthroscopic evidence of articular cartilage degeneration. Arthritis Rheum. Wiley Online Library; 1991;34(11):1381–1386.
\newline
13. Yang X, Li Z, Cao Y, et al. Efficacy of magnetic resonance imaging with an SPGR sequence for the early evaluation of knee cartilage degeneration and the relationship between cartilage and other tissues. J Orthop Surg Res. BioMed Central; 2019;14(1):152.
\newline
14. Hong Z, Chen J, Zhang S, et al. Intra-articular injection of autologous adipose-derived stromal vascular fractions for knee osteoarthritis: a double-blind randomized self-controlled trial. Int Orthop. Springer; 2019;43(5):1123–1134.
\newline
15. Peterfy CG, Guermazi A, Zaim S, et al. Whole-organ magnetic resonance imaging score (WORMS) of the knee in osteoarthritis. Osteoarthr Cartil. Elsevier; 2004;12(3):177–190.
\newline
16. Kretzschmar M, Lin W, Nardo L, et al. Association of physical activity measured by accelerometer, knee joint abnormalities, and cartilage T2 measurements obtained from 3T magnetic resonance imaging: data from the Osteoarthritis Initiative. Arthritis Care Res (Hoboken). Wiley Online Library; 2015;67(9):1272–1280.
\newline
17. Roemer FW, Frobell R, Lohmander LS, Niu J, Guermazi A. Anterior Cruciate Ligament OsteoArthritis Score (ACLOAS): longitudinal MRI-based whole joint assessment of anterior cruciate ligament injury. Osteoarthr Cartil. Elsevier; 2014;22(5):668–682.
\newline
18. Crawford R, Walley G, Bridgman S, Maffulli N. Magnetic resonance imaging versus arthroscopy in the diagnosis of knee pathology, concentrating on meniscal lesions and ACL tears: a systematic review. Br Med Bull. Oxford University Press; 2007;84(1):5–23.
\newline
19. Bien N, Rajpurkar P, Ball RL, et al. Deep-learning-assisted diagnosis for knee magnetic resonance imaging: development and retrospective validation of MRNet. PLoS Med. Public Library of Science; 2018;15(11):e1002699.
\newline
20. Liu F, Guan B, Zhou Z, et al. Fully Automated Diagnosis of Anterior Cruciate Ligament Tears on Knee MR Images by Using Deep Learning. Radiol Artif Intell. Radiological Society of North America; 2019;1(3):180091.
\newline
21. Chang PD, Wong TT, Rasiej MJ. Deep Learning for Detection of Complete Anterior Cruciate Ligament Tear. J Digit Imaging. Springer; 2019;1–7.
\newline
22. Germann C, Marbach G, Civardi F, et al. Deep Convolutional Neural Network–Based Diagnosis of Anterior Cruciate Ligament Tears: Performance Comparison of Homogenous Versus Heterogeneous Knee MRI Cohorts With Different Pulse Sequence Protocols and 1.5-T and 3-T Magnetic Field Strengths. Invest Radiol. 9000.
\newline
23. Cui Y, Song Y, Sun C, Howard A, Belongie S. Large scale fine-grained categorization and domain-specific transfer learning. Proc IEEE Conf Comput Vis pattern Recognit. 2018. p. 4109–4118.
\newline
24. Anumanchipalli GK, Chartier J, Chang EF. Speech synthesis from neural decoding of spoken sentences. Nature. Nature Publishing Group; 2019;568(7753):493.
\newline
25. Krizhevsky A, Sutskever I, Hinton GE. Imagenet classification with deep convolutional neural networks. Adv Neural Inf Process Syst. 2012. p. 1097–1105.
\newline
26. Warrens MJ. Cohen’s linearly weighted kappa is a weighted average. Adv Data Anal Classif. Springer; 2012;6(1):67–79.
\newline
27. Zhou X-H, McClish DK, Obuchowski NA. Statistical methods in diagnostic medicine. John Wiley $\&$  Sons; 2009.
\newline
28. Han S, Pool J, Tran J, Dally W. Learning both weights and connections for efficient neural network. Adv Neural Inf Process Syst. 2015. p. 1135–1143.
\newline
29. Huang G, Liu Z, Van Der Maaten L, Weinberger KQ. Densely connected convolutional networks. Proc IEEE Conf Comput Vis pattern Recognit. 2017. p. 4700–4708.
\newline
30. Redmon J, Divvala S, Girshick R, Farhadi A. You only look once: Unified, real-time object detection. Proc IEEE Conf Comput Vis pattern Recognit. 2016. p. 779–788.
\newline
31. Milletari F, Navab N, Ahmadi S-A. V-net: Fully convolutional neural networks for volumetric medical image segmentation. 2016 Fourth Int Conf 3D Vis. IEEE; 2016. p. 565–571.

\newpage
\section{Appendix E1}
Two VNet architectures localized the volume of knee that included the ACL; each VNet consisted of four down-sampling convolutions to learn an encoding for the volume, followed by four upward-convolutions to decode the segmentation mask (31). The VNet segmentation is a portion of a larger project aimed towards segmenting all knee tissues, which we used to isolate the ACL region of interest.

The first VNet segmented the patellar cartilage, femur, tibia, and meniscus from the image background. These five compartments were further segmented by a second VNet into 11 compartments consisting of patellar cartilage, medial and lateral femoral condyles, medial and lateral tibial cartilage, tibia, four meniscal horns, and background.

The VNets were trained and tested on a total of 480 manual segmentations. Rotations between -5 and 5 degrees were applied to augment the segmentations. All parameters of the originally published VNet were maintained with the exception of the residual connections (31). The VNet was trained for 100 epochs using the following parameters: 10\textsuperscript{-5} learning rate, batch size of 1, Adam optimizer, with combined weighted cross entropy and Dice loss functions. This combined loss function enables increased penalization when the meniscus or cartilage are misclassified as background, mitigating the relatively small proportions of these two compartments in the entire knee.

To avoid any additional manual annotation, the femoral and tibial cartilage masks generated by the concatenated VNets were then used to create bounds on the original knee MRI and isolate the ACL. Specifically, the ACL’s superior bound was determined by the superior point of the femoral condyle, inferior bound by inferior point of tibial cartilage, anterior bound by anterior point of tibial cartilage, posterior bound by posterior point of femoral cartilage, medial bound by medial point of femoral condyle, and lateral bound by medial point of lateral femoral condyle. All bounded ligament volumes were resized to the mean calculated volume of 98 x 82 x 47 pixels, and left knees were axially mirrored to match right knee orientation. We further calculated the intersection over union between manually segmented and automated VNet bound ACL regions in order to quantify the metric of overlap between clinical and deep learning segmentation.
\end{document}